%% file: zz_Manuscript.tex
\documentclass[
reprint,
amsmath,amssymb,aps,prl]{revtex4-2}
\usepackage{xcolor}
\usepackage{graphicx}
\usepackage{dcolumn}
\usepackage{bm}
\usepackage{animate}

\newcommand\Alfven{Alfv\'en } 
\newcommand\Alfvenic{Alfv\'enic } 

\begin{document}


\title{How the Oblique Drift Instability Alters Solar Wind Heating and Constrains the Distribution of Solar Wind Observations}


\author{Mihailo M. Martinovi\'c}
\email{mmartinovic@arizona.edu}
\affiliation{Lunar and Planetary Laboratory, University of Arizona, Tucson, AZ 85721, USA.}
  
\author{Kristopher G. Klein}
\affiliation{Lunar and Planetary Laboratory, University of Arizona, Tucson, AZ 85721, USA.}

\author{Leon Ofman}
\affiliation{The Catholic University of America, Washington, DC 20064, USA} 
\affiliation{Heliophysics Science Division, NASA Goddard Space Flight Center, Greenbelt, MD 20771, USA} 
\affiliation{Visiting, Tel Aviv University, Tel Aviv, Israel}

\author{Yogesh}
\affiliation{Department of Physics and Astronomy, University of Iowa, IA 52242, USA}

\author{Jaye L. Verniero}
\affiliation{Heliophysics Science Division, NASA, Goddard Space Flight Center, Greenbelt, MD 20771, USA}

\author{Peter H. Yoon} 
\affiliation{Institute for Physical Science and Technology, University of Maryland, College Park, MD 20742-2431, USA} 

\author{Gregory G. Howes}
\affiliation{Department of Physics and Astronomy, University of Iowa, IA 52242, USA}

\author{Daniel Verscharen}
\affiliation{Mullard Space Science Laboratory, University College London, Dorking, RH5 6NT, UK} 

\author{Benjamin L. Alterman}
\affiliation{Heliophysics Science Division, NASA Goddard Space Flight Center, Greenbelt, MD 20771, USA} 

\date{\today}

\begin{abstract}
Ion-driven plasma instability thresholds, derived from linear theory, constrain the distribution of solar observations in parameter space, defining boundaries of stable plasma parameters.
Excursions beyond these thresholds result in the emission of energy, transferred from particles to coherent electromagnetic waves, acting to adjust the system toward a more stable configuration.
In this work, we use linear Vlasov--Maxwell theory to define parametric limits for a low-$\beta$ plasma that contains a drifting proton beam or helium ($\alpha$-particle) population. 
A sufficiently fast and dense drifting population triggers an Oblique Drift Instability (ODI). 
This instability decreases the velocity drift between the thermal core proton and secondary populations and prevents the ratio of core thermal to magnetic pressure $\beta_c$ from decreasing below a minimum value by increasing the temperatures---i.e. heating---of both the core and drifting populations. 
Our theoretical results are of interest for \emph{Parker Solar Probe} observations, as they provide an additional mechanism for perpendicular heating of ions active in the sub-\Alfvenic solar wind. 
The ODI may explain the discrepancy between long-standing expectations of measurements of very low-$\beta$ plasmas with very large ion temperature anisotropies in the near-Sun environment and \emph{in situ} observations, where $\beta$ is consistently measured above a few percent and the secondary ion populations drift faster than the bulk of proton population by no more than approximately the local \Alfven speed. 
\end{abstract}

\maketitle



\paragraph {\textbf{Introduction}.}

\begin{figure}
\includegraphics[width=0.48\textwidth]{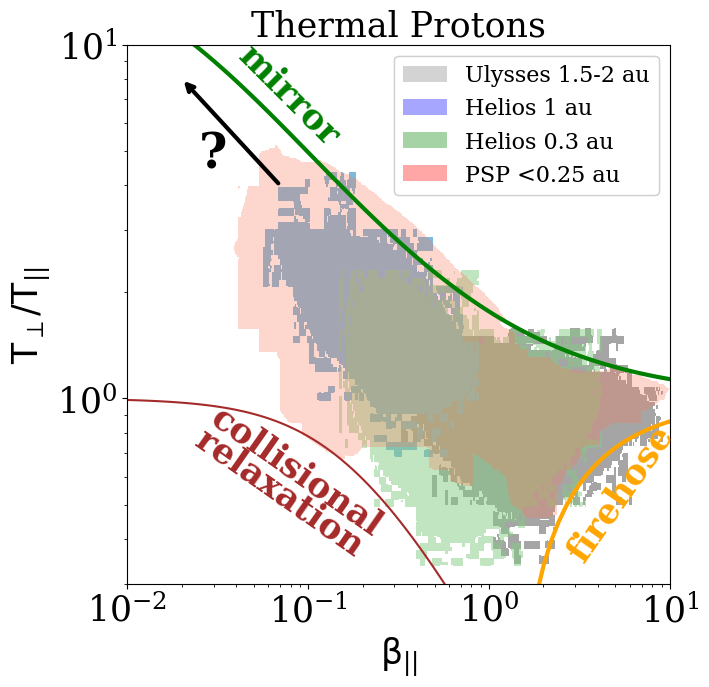}
\caption{\label{fig:brazil} Summary of proton measurements of previous missions. 
The three sides of the plot limit the ``allowed'' parameter space, while the plasma moving to lower $\beta$ state does not have obvious limitations in linear theory of single population. 
}
\end{figure}

Many astrophysical plasmas, including the solar wind in the inner heliosphere, are weakly collisional, allowing departure from Local Thermodynamic Equilibrium (LTE). 
Characterizing the thermodynamics of weakly collisional plasmas is a long-standing open question, whose answer is central to explaining the heating and acceleration of the solar wind \cite{Marsch_2012_SSRv,Verscharen_2019_LRSP}. 
For decades, surveys of in situ solar wind proton observations have been routinely represented with the so-called ``Brazil'' plot---a phase space determined by $\beta_{\parallel,p}$---thermal to magnetic pressure ratio of protons---and temperature anisotropy $T_{\perp,p} / T_{\parallel,p}$, with $\perp$ and $\parallel$ defined with respect to the background magnetic field vector $\mathbf{B}$ (Fig. \ref{fig:brazil}). 
The ``Brazil'' plot  demonstrates that the solar wind evolution is limited by instabilities \cite{Matteini_2013_JGRA}, where in high-$\beta$ wind, usually further away from the Sun, even the smallest changes in anisotropy can cause the thermal-pressure-dominated plasma to become intensely unstable \cite{Arzamasskiy_2023_PhRvX}. 
The data is remarkably well constrained by the parametric theoretical thresholds \cite{Bale_2009_PRL,Verscharen_2016_ApJ} of the ion-cyclotron (IC) and mirror instabilities for $T_{\perp,p} / T_{\parallel,p} > 1$ and the oblique and parallel firehose instabilities for $T_{\perp,p} / T_{\parallel,p}<1$ \cite{Yoon_2017_RvMPP,Verscharen_2019_LRSP}.
On the contrary, the low $\beta$ region, constrained by collisonal relaxation for $T_{\perp,p} / T_{\parallel,p} < 1$, has an ``open-ended'' part, where even strongly anisotropic plasma can remain linearly stable. 
Linear theory for a single ion-component does not provide any limit on how low plasma $\beta_{\parallel,p}$ can be. 
However, for $T_{\perp,p} / T_{\parallel,p} > 1$, the observations of $\beta_{\parallel,p}<0.1$ are relatively rare beyond 0.3 AU \cite{Maruca_2023_A&A,Yogesh_2025_arxiv}. 
Due to the lack of \emph{in situ} observations, low-$\beta$ instability thresholds for inner heliospheric conditions have received some attention \cite{Vafin:2019,Yoon_2024_ApJ_Coll_P}, but not been investigated as rigorously as those with $\beta \sim 1$. 
We propose an instability mechanism driven by relative drifts between ion populations that limits the solar wind from filling the low-$\beta$, high $T_{\perp,p}/T_{\parallel,p}$ region of parameter space.

Launched in 2018, \emph{Parker Solar Probe (PSP)} \cite{Fox_2016} has reached the so-called \Alfven surface \cite{Kasper_2021_PhRvL}, inside which the solar wind bulk speed is lower than the \Alfven speed $v_A = w_{\parallel,p} / \beta_{\parallel,p}^{1/2}=B/\sqrt{\mu_0 m_p n_p}$, where $w_{\parallel,p}^2 = 2 k_B T_{\parallel,p} / m_p$ is the proton thermal speed and $m_p$ is the mass of the proton. 
We define parallel proton plasma beta $\beta_{\parallel,p}= 2 \mu_0 n_p k_b T_{\parallel,p}/B^2$ and temperature anisotropy $T_{\perp,p} / T_{\parallel,p}$;  $\mu_0$, $k_b$, and $n_p$ are the magnetic permeability of vacuum, Boltzmann constant, and proton density, respectively.
At distances within the Alfven surface, \Alfven waves generated in situ can travel back to the surface of the Sun.  
It has been hypothesized that inside this surface the solar wind heating is significantly enhanced \cite{Meyer-Vernet_2007,Kasper_2019_Alfven}. 
Various mechanisms have been invoked to explain preferential perpendicular heating in this region, e.g. Stochastic Heating \cite{Chandran_2010,Martinovic_2020_ApJS} or IC Heating \cite{Hollweg_1999b,Cranmer_2000,Bowen_2024_ApJ}. 
Initial in situ measurements in a sub-\Alfvenic plasma \cite{Bowen:2025,Adhikari:2025} have not consistently shown the enhancement of these processes in the near-Sun solar wind at or below the \Alfven surface.

Each component of the plasma velocity distribution function (VDF) can be anisotropic, with distinct temperatures perpendicular and parallel to the direction of the magnetic field \cite{Matteini_2007_GRL}. 
The two populations can also have different speeds along the magnetic field, producing a commonly observed inter-component drift \cite{Durovcova_2019_SoPh}.
These anisotropies and relative drifts often store a significant percentage of the system's ``free'' energy \cite{Chen_2016_ApJ,Klein_2018,Martinovic_2023_ApJ_Ins_2}. 
Significant departures from LTE can drive the system unstable, generating electromagnetic fluctuations that act to reduce the free energy source of the instability \cite{Stix_1992,Gary_1993}. 
The predictions made by linear Vlasov--Maxwell theory imply that  the onset of instabilities at kinetic scales effectively constrain the non-thermal features of the VDF and, consequently, the parameter space distribution
\cite{Kasper_2002_GRL,Matteini_2007_GRL,Huang_2020_ApJS,Martinovic_2021_ApJ_Ins_1,Bandyopadhyay_2022_PhPl,Tong:2019,Verscharen:2022}. 

We hypothesize that this low-$\beta$ limit is evidence that an Oblique Drift Instability (ODI) is converting free energy carried by the relative drift between $\alpha$-particles and/or a proton beam (with respect to the proton core) into both perpendicular and parallel thermal energy.
This heating does not require the interaction of particles with the background turbulence, nor with existing coherent waves. 
The process is driven by the same type of wave-particle resonance and only slightly quantitatively different when driven by drifting protons or $\alpha$-particles.  
We find that the ODI is important for both limiting the low-$\beta$ values by heating the protons and for constraining $\Delta v_\alpha$ to be consistently below the local \Alfven speed \cite{Wang_2025_ApJ}.


\paragraph {\textbf{Linear Theory of the low-$\beta$ ODI}.}
Assuming a combination of relatively drifting bi-Maxwellian VDFs, the solution to the wave equation $|\underline{\mathbf{D}} \big( \omega, \mathbf{k}, \mathcal{P} \big)|=0$ for a hot, magnetized plasma allows an estimate of the complex frequency at maximum wave growth rate $\omega=\omega_{\textrm{r}} + i \gamma_{\mathrm{max}}$ associated with wavevector $\mathbf{k}$ and plasma parameters $\mathcal{P}$, the power emitted for each component over the course of a single wave period $P_j$, electric $\mathbf{\delta E}$ and magnetic $\mathbf{\delta B}$ field fluctuations and polarizations for the Most Unstable mode (MUM). 
Here, we use the \texttt{PLUME} and \texttt{PLUMAGE} solvers when a bi-Maxwellian VDF shape is assumed \cite{Klein_2015_PhPl,Klein_2017_JGRA,Klein_2025_RNAAS} and \texttt{ALPS} solver \cite{Verscharen_2018_JPlPh,Klein_2025_PhPl} to analyze the modes driven by non-Maxwellian VDFs obtained from simulations. 

\begin{figure}[b]
\includegraphics[width=0.48\textwidth]{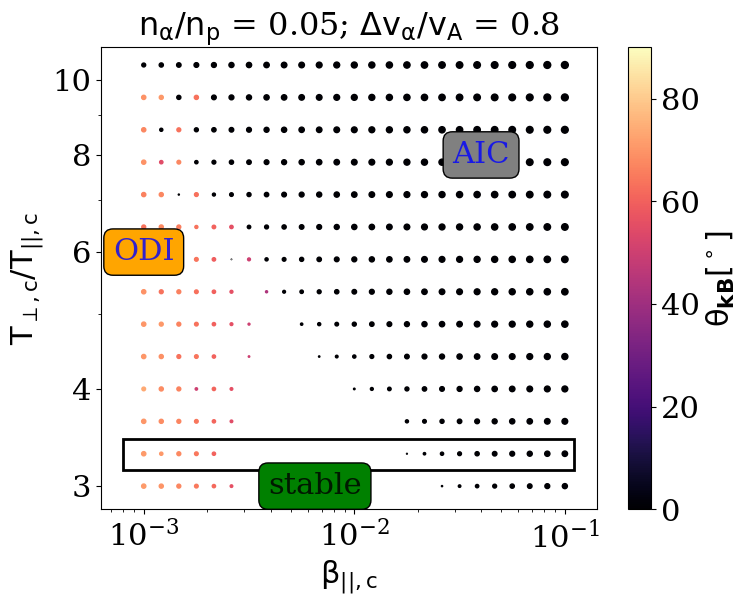}
\caption{\label{fig:overview} The ``island'' of stability between the forward \Alfven mode unstable in the direction parallel to $\mathbf{B}$--- the proton AIC instability--- and the Oblique Drifting Instability (ODI) at lower proton core $\beta_c$ for fixed $n_\alpha$ and $\Delta v_{\alpha}$ 
The size of points increases as $\lg(\gamma_{\mathrm{max}}/\Omega_p)$, covering the range between $10^{-3}$ and $0.3$. 
}
\end{figure}

\begin{figure}
\includegraphics[width=0.48\textwidth]{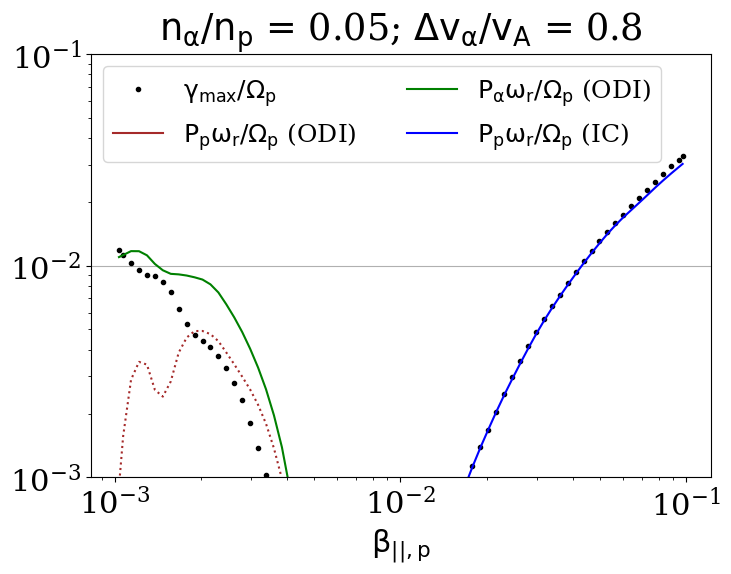}
\caption{\label{fig:line_test} Overview of the ion contributions to the AIC and ODI instabilities for intervals marked with black rectangle on Fig. \ref{fig:overview}, with solid/dashed lines indicating power emission/absorption.
Both instabilities emit a forward-propagating \Alfven mode. 
\texttt{PLUMAGE} growth rate and \texttt{PLUME} power transmission solutions marked with dots and lines, respectively. 
Note that in the linear approximation $\sum_j P_j \approx \gamma / \omega_r$ \cite{Stix_1992}. 
}
\end{figure}

Fig. \ref{fig:overview} shows \texttt{PLUMAGE} analysis of core proton and drifting $\alpha$-particle component for a set of 625 logarithmically spaced $\beta_{\parallel,p}$--$T_{\perp,p}/T_{\parallel,p}$ pairs of values with all the plasma parameters $\mathcal{P}$ ($n_\alpha/n_p$, $\Delta v_\alpha/v_A$, $\Delta v_\alpha/v_A$, $T_{\perp,p,\alpha}/T_{\parallel,p,\alpha}$) fixed, colored by the MUM angle of propagation $\theta_{\mathbf{kB}}$. 
We set $T_{p,\perp} / T_{p,\parallel}=3$, and $T_{\perp,\alpha}/T_{\parallel,\alpha} = T_{\parallel,\alpha}/T_{\parallel,p} = 1$ to focus on the drift as the main source of free energy. 
An oblique drift-induced mode will become unstable at low-$\beta$ with both $\gamma_{\mathrm{max}}/\Omega_p$ and the angle of propagation $\theta_{\mathbf{kB}}$ increasing as $\beta_{\parallel,p}$ decreases. 
We note the existence of an ``island'' of stability for $\Delta v_{\alpha}/v_A < 0.9$, $\beta_{\parallel,p} \in [0.005,0.02]$, and $T_{\perp,p} / T_{\parallel,p} \lesssim 5$---conditions common for near-Sun solar wind \cite{Wang_2025_ApJ}.

We further investigate a slice of constant proton anisotropy, denoted with a \emph{black rectangle} in Fig. \ref{fig:overview}, using \texttt{PLUME} to identify nine solutions: $\alpha$-cyclotron, Alfv\'en, fast magnetosonic, and slow magnetosonic waves---all four having both distinct forward and backward propagating solutions in the plasma frame due to velocity space symmetry breaking associated with the drifting components, and a non-propagating entropy mode. 
In Fig.~\ref{fig:line_test} \emph{black dots} illustrate the \texttt{PLUMAGE} inferred MUM growth rate and \texttt{PLUME} inferred emitted (absorbed) power solutions for both parallel AIC and oblique ODI, both of which originate from the same forward propagating \Alfven mode. 
For drifts lower than the \Alfven speed, we see the range of $\beta_{\parallel,p}$ values for which the plasma is stable, with lower- and higher-$\beta$ cases dominated by the oblique and parallel unstable modes, respectively. 
If the drift is increased to $\Delta v_\alpha/v_A = 1$, both modes can be active in the same interval, implying that the oblique waves can still extract energy from $\alpha$ component while the intensively growing parallel AIC mode is active, as will be demonstrated in nonlinear hybrid simulations. 


\begin{figure}
\includegraphics[width=0.47\textwidth]{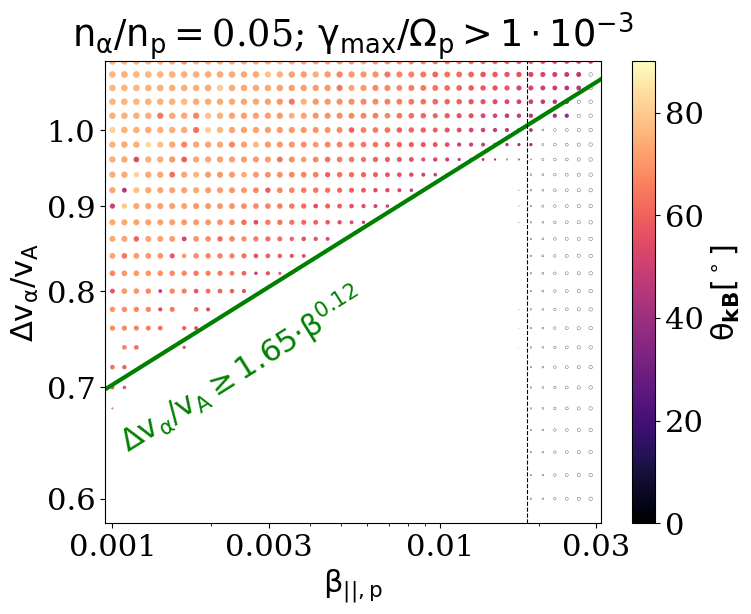}
\caption{ODI mode growth rates and propagation angles (as in Fig. \ref{fig:overview}) dependence on the velocity drift.
The best-fit parametric instability threshold is shown in green.
}
\label{fig:beta_vva_threshold}
\end{figure}

For a closer examination of the ODI, we show its emitted power as a function of $\Delta v_{\alpha} / v_A$. 
Crucially, the instability threshold for the ODI requires smaller $\alpha$-particle drifts to overcome the proton emission as $\beta_\parallel$ decreases. 
Plasmas with $\beta_\parallel < 1\%$ have a maximum stable value of $\Delta v_{\alpha} / v_A$ below 0.85, implying that this type of resonant instability will become active prior to any firehose-like instability where excess parallel pressure is the main free energy source \cite{Chen_2016_ApJ}. 
The mode properties and the maximum drifts are preserved for low $\alpha$-particle densities and only marginally vary with $n_\alpha/n_p$, while not varying with proton parameters (not shown). 

\begin{figure}
\includegraphics[width=0.47\textwidth]{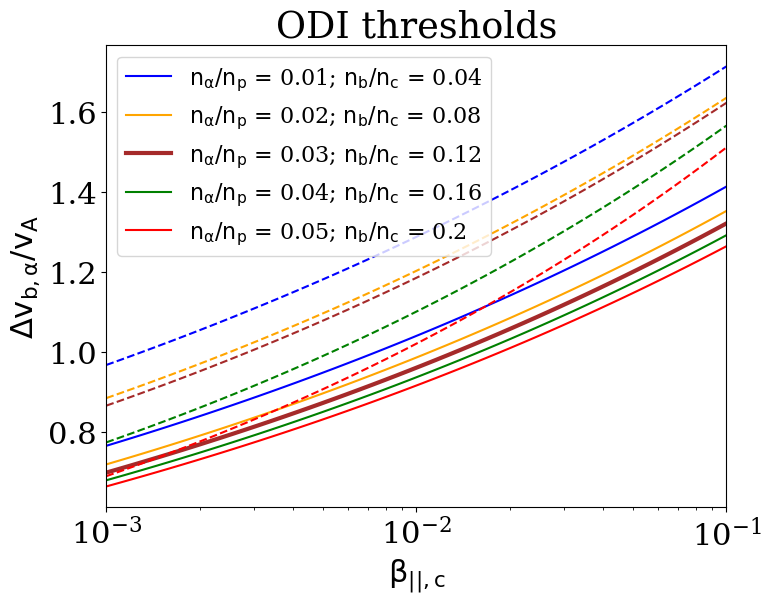}
\caption{ODI instability thresholds for varying $\alpha$-particle densities (\emph{solid lines}) for $\gamma/\Omega_p=10^{-3}$. 
\emph{Dashed lines} provide comparison given by \citet{Gary_1993} for the proton-beam induced ODI with beam densities $n_b$ with the same mass flux as the $\alpha$ cases. 
}
\label{fig:numerical_thresholds}
\end{figure}

Fig.~\ref{fig:beta_vva_threshold} illustrates that the ODI instability threshold in logarithmic space very nearly follows a straight line. 
This feature holds at least up to $n_\alpha/n_p \lesssim 10\%$ and for growth rates ranging from $\gamma_{\mathrm{max}}/\Omega_p \in (10^{-4}, 2 \cdot 10^{-2})$. 
This trend enables a linear fit of $\Delta v_\alpha / v_A = \hat{m} \log_{10}(\beta_{\parallel,p}) + \hat{b}$ to obtain the numerical expression for the threshold, with the example labeled along the fitted \emph{green} line. 
We repeat the fitting procedure for various density ratios. 
Fitting slopes and intercepts for different $n_\alpha/n_p$ while keeping the growth rate fixed enables another stage of linear fits with relative density as the independent variable. 
These results can be combined to obtain the instability condition for any onset growth rate. 
For $\gamma_{\mathrm{max}}/\Omega_p = 10^{-3}$ we get 

\begin{equation}
    \Delta v_\alpha / v_A \geq 10^{0.255 - 0.4 n_\alpha/n_p} \beta_{\parallel,p}^{0.116 - 0.38 n_\alpha/n_p}
\end{equation}
This analytical representation accurately describes the $\alpha$-particle-induced ODI growth rates shown in Fig. \ref{fig:numerical_thresholds}.

We directly compare this instability to the one described in Section 8.5 of \citet{Gary_1993}, where the proton beam drives the ODI with the drift being the primary source of free energy. 
We expand on the results of \cite{Gary_1993} to normalize relative proton beam density in such a way that the mass flux of \emph{proton} beams and $\alpha$-particles remains equal, showing them in the same color on Fig. \ref{fig:numerical_thresholds}. 
The drift value required for $\alpha$-particles to drive the instability is consistently below the one for proton beam over a wide range of $\beta$ values due to difference in resonance conditions for the underlying $n=-1$ resonance (see next Section). 
As the proton beams are not consistently present in the solar wind \cite{Alterman_2019_PhDT,Verniero_2020_ApJS,Short_2024_ApJL}, $\alpha$-particles remain the primary driver of the ODI and regulator of the low-$\beta$ limit in the solar wind. 

\paragraph {\textbf{Plasma $\beta$ Limits in Hybrid Simulations}}
To illustrate the transfer of energy from drifting $\alpha$-particles to thermal protons via the ODI, we use a well-established 2.5D hybrid-PIC code that captures ion dynamics across multiple ion-cyclotron periods with realistic proton mass and Alfv\'{e}n-to-light speed ratio, bypassing standard PIC code limitations by representing electrons as a massless neutralizing fluid \cite{Ofman_2007_JGRA,Ofman_2010_JGRA,Ofman_2019_JGRA}. 
Our setup is similar to recent studies \cite{Ofman_2022_ApJ,Ofman_2023_ApJ,Ofman_2025_ApJ}---we use a $256^2$ spatial grid with 512 Particles Per Cell (PPC), fully sufficient to capture plasma dispersion at ion scales. 
Full details for the simulation can be found in the supplemental information appendix.

\begin{figure}
    \centering
    \includegraphics[width=1.0\linewidth]{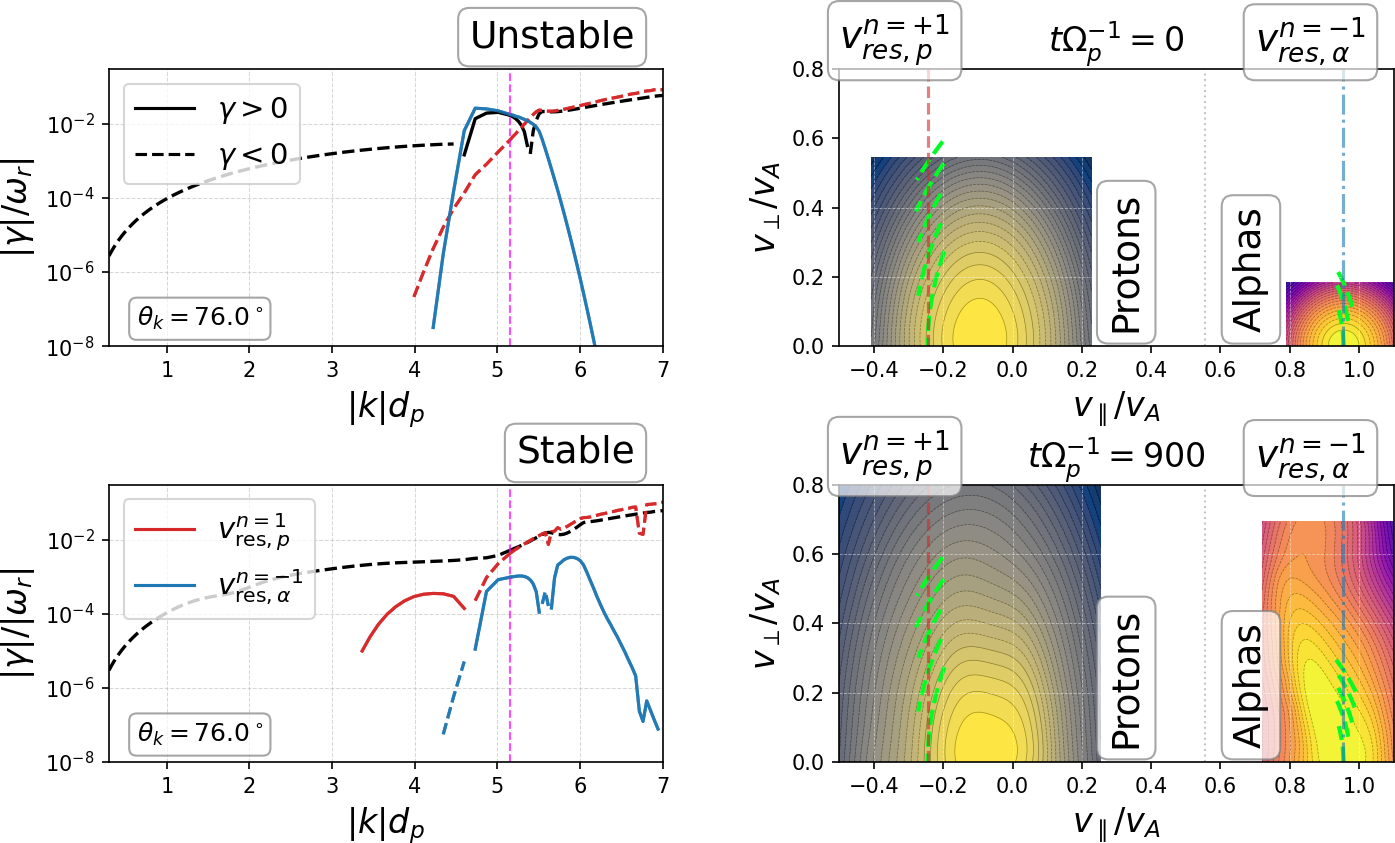}
    \caption{Growth (solid lines, left column) and damping rates (dashed) for proton and $\alpha$ VDFs representative of the young solar wind unstable to the ODI at early (top right) and late (bottom right) times in the simulation. 
    The emitting $\alpha$-particle $(n=-1)$ and absorbing proton $(n=1)$ cyclotron resonances for the most unstable mode are marked by blue and red lines. 
    Particles are scattered along resonance curves (green) for the highlighted wavevector (magenta, left column), leading to wave emission and heating. }
    \label{fig:alps}
\end{figure}

The \emph{top} panels of Fig.~\ref{fig:alps} show the initial state of the simulation. 
The $\alpha$-particles of density $n_\alpha/n_p = 5\%$ drift at $\Delta v_\alpha/v_A = 0.95$ with respect to protons, driving the ODI for the low $\beta_{\parallel,p} = 0.003$ conditions. 
The initial values of $T_{p,\perp} / T_{p,\parallel}=3$ and $T_{\alpha,\perp} / T_{\alpha,\parallel}=1$ are stable with respect to AIC. 
The ODI is induced by the drift of secondary populations---proton beams and alpha-particles---with the wavevector of the unstable mode at an oblique angle with respect to the background magnetic field $\mathbf{B}_0$. 
The ODI couples via cyclotron resonances, where the relation of the resonant particle velocity $\mathbf{v_{res}}$ (\emph{green dashed} lines), real frequency $\omega_r$ and wavenumber $\mathbf{k}$ is given as $\omega_r - k \cdot \mathbf{v_{res}} - n\Omega_s = 0$, where $\Omega_s$ is the cyclotron frequency for species $s$ and $n$ is an integer. 
At wavelengths a fraction of the inertial length, but notably larger than the gyroradius, the $n=-1$ cyclotron resonance (\emph{blue} line) emits an unstable mode, scattering initially isotropic $\alpha$-particles along the \emph{green} resonance curves in the velocity space. 
These linear dispersion relations are calculated directly from the simulated VDFs using \texttt{ALPS} \cite{Verscharen_2018_JPlPh,Klein_2025_PhPl}; see supplemental information for more details.
The emitting wavevectors are shown by \emph{solid} lines in the \emph{left} panels, while non-negligible damping is marked by \emph{dashed} lines, showing notable absorption by protons through the $n=1$ resonance, marked in \emph{red}.

During later stages of the simulation (\emph{bottom} panels in Fig.~\ref{fig:alps}), some of the free energy from the $\alpha$-particle drift is drained, decreasing $\Delta v_\alpha/v_A$, but increasing $T_{\perp,\alpha}$, $T_{\perp,p}$ and $T_{\parallel,p}$ in the process. 
This final state is stable with respect to ODI, as the emission of $\alpha$-particles does not surpass the absorption of the protons (\emph{bottom left}), but the transfer of energy from one component to another is still active, and the protons continue to be heated via $n=-1$ $\alpha$-particle resonance (magenta line). 

This implies a prolonged process that will remain active until the free energy from the velocity drift is drained, heating both the proton and $\alpha$-particle distributions in both parallel and perpendicular directions. 
The increase in $T_{p,\perp}$ leads to a secondary intense forward propagating AIC parallel propagating instability, depositing some of the drift free energy into coherent plasma waves. 
The fraction of energy deposited in the parallel direction increases $T_{p,\parallel}$ and thus $\beta_{p,\parallel}$. 
This process is more intensive for lower $\beta_{p,\parallel}$, hence putting a limit to its minimum value. 


\paragraph{\textbf{Discussion and Conclusions.}}
The AIC and mirror instability provide a threshold for the maximum value of proton anisotropy in the solar wind at $\beta \gtrsim 1$ \cite{Kasper_2002_GRL,Bale_2009_PRL}. 
While there are quasi-linear models that hypothesize the low-$\beta$, $T_{\perp,p} / T_{\parallel,p}<1$ region is regulated by collisions \cite{Vafin:2019,Yoon_2024_ApJ_Coll_P}, the low-$\beta$, $T_{\perp,p} / T_{\parallel,p}>1$ region has no solid theoretical nor observational constraints. 
However, the measurements obtained from \emph{PSP} encounters (perihelia) shows higher lower-bound values of $\beta$ compared to pre-launch predictions \cite{Bale_2016}. 

We propose an instability mechanism---the Oblique Drift Instability---to explain the lower limit on the observations of $\beta_{\parallel,p}$, the magnitudes of $\alpha$-particle drifts with respect to the protons, which are also very rarely observed to be faster than a single \Alfven velocity \cite{McManus_2024_ApJ,Wang_2025_ApJ}, as well as the heating of both the protons and alphas. 
The presence of drifting $\alpha$-particles leads to emission of power for oblique wavemodes via the $n=-1$ cyclotron resonance. 
If $\beta_{\parallel,p}$ is sufficiently large, the increased width of the proton VDF fills the velocity space with enough particles that can absorb this emitted power, creating a stable configuration. 
In contrast, lowering the value of $\beta_{\parallel,p}$ decreases the proton phase space density at the appropriate resonant velocities, enabling the $\alpha$-driven emission to destabilize the equilibrium. 
The emitted power from the alphas is then absorbed by thermal protons through the $n=1$ cyclotron resonance, leading to proton heating and increasing $\beta_{\parallel,p}$. 
This setup effectively restricts the lowest values of $\beta_{\parallel,p}$ ``allowed'' by the linear theory. 
It also provides a constant mechanism to both limit the secondary population drift and to heat thermal protons and secondary populations.

As the set of VDF parameters specific for ODI correspond to the ones observed in the sub-\Alfvenic solar wind, 
the ODI may serve as an additional avenue of intense solar wind heating in this region of the inner heliosphere. 
Just outside of the corona ($r \sim 2 R_\odot$) the role of ODI in unclear. 
Although \Alfven speed is very high compared to the thermal speed, it is not currently understood how the $\alpha$-particle and proton beam drifts evolve once the collisions are no longer dominant. 
At larger radial distances plasma $\beta$ becomes too high for the ODI to be significant, leaving the sub-\Alfvenic wind as a specific region where this mechanism can become very intense and act as the primary regulator of solar wind evolution. 

This result will be further tested with \emph{PSP} solar encounter observations as statistically significant amounts of observations in low-$\beta$ solar wind are being collected by the spacecraft. 
The caveats of the observational work will be addressed in future work, as the data products of the mission mature and are able to confidently separate multiple proton components as well as a potential $\alpha$-particle beam \cite{McManus_2024_ApJ,Martinovic_2025_ApJL}. 
Finally, since the instability thresholds can be very steep in velocity space, the future work requires careful treatment of instrument uncertainties when comparing measured VDFs with our predictions.


\begin{acknowledgments}
M. M. Martinovi\'c and K. G. Klein were financially supported by NASA grants: 80NSSC22K1011, 80NSSC19K1390, 80NSSC23K0693, 80NSSC19K0829, 80NSSC24K0724. 
An allocation of computer time from the UA Research Computing High Performance Computing at the University of Arizona is gratefully acknowledged. L. Ofman and Y. acknowledge support by NASA grant 80NSSC24K0724 and NSF SHINE grant AGS-2300961. 
Yogesh acknowledges the support by the College of Liberal Arts and Sciences at the University of Iowa.
D. Verscharen is supported by STFC Consolidated Grant ST/W001004/1. 
B. L. Alterman is supported by NASA grant 80NSSC22K1011. 
The authors would also like to thank members of ISSI International Team \#563 supported by the International Space Science Institute (ISSI) in Bern for productive conversations regarding this work. 
\end{acknowledgments}

\bibliography{Latex_Refs,additional_ref}


\clearpage
\input{SI.tex}

\end{document}

%% file: SI.tex

\renewcommand{\thepage}{S\arabic{page}}
\setcounter{page}{1}

\renewcommand{\thesection}{S\arabic{section}}
\setcounter{section}{0}

\renewcommand{\thefigure}{S\arabic{figure}}
\setcounter{figure}{0}

\renewcommand{\thetable}{S\arabic{table}}
\setcounter{table}{0}

\begin{center}

\emph{Physical Review Letters}\\

    Supporting Information for\\

\textbf{How the Oblique Drift Instability Alters Solar Wind Heating and Constrains the Distribution of Solar Wind Observations}

\end{center}

\section{Numerical Evaluation of Bi-Maxwellian Linear Stability}
\label{sec:si.plume}

The plasma stability in this article is predicted using two types of underlying assumptions. 
First, for the theoretical threshold calculations, we assume that the VDF is consisted of components that are bi-Maxwellian, with each component having different temperatures in parallel $T_{\parallel}$ and perpendicular $T_\perp$ directions with respect to the background magnetic field. 
While analyzing the results of hybrid simulations, the assumption of Maxwellianity is dropped and the VDFs are accessed as calculated by the simulation. 

When the VDF is represented with bi-Maxwellian components, the dispersion relation can be calculated for a given set of global dimensionless parameters
\begin{equation}
  \mathcal{P}_{0}=\left(\beta_{\parallel,p},\frac{w_{\parallel,p}}{c}
  \right)
\end{equation}
with $c$ being the speed of light, and parameters specific to component $j$ 
\begin{eqnarray}
  \mathcal{P}_{j}=
  \left(\frac{n_j}{n_p},\frac{T_{\perp,j}}{T_{\parallel,j}},
  \frac{T_{\parallel,j}}{T_{\parallel,p}}, \frac{\Delta
    v_{j}}{v_{Ap}}, \frac{m_j}{m_p}, \frac{q_j}{q_p} \right);
\end{eqnarray}
see Chapter 10 of Stix 1992 \cite{Stix_1992} for the full set of equations. 
The Plasma in a Linear Uniform Magnetized Environment (\texttt{PLUME}) code \cite{Klein_2015_PhPl,Klein_2025_RNAAS}, which solves the linear dispersion relation for the hot Vlasov-Maxwell system of equations, is used to locate the complex frequencies and associated eigenfluctuations of the supported linear modes. 

A significant upgrade to this traditional approach is given by \texttt{PLUMAGE} \cite{Klein_2017_JGRA}, which performs a Nyquist integral over the $\gamma > 0$ half-complex frequency plane, diagnosing the existence of any unstable modes in the system without the necessity to locate and investigate each mode solution individually. 
This approach can be implemented automatically, requiring lower CPU resources and enabling the investigation of large number of VDFs \cite{Klein_2019_ApJ,Martinovic_2021_ApJ_Ins_1}. 
\texttt{PLUMAGE} provides solutions for $\omega_{\mathrm{max}} = \omega_{r,\mathrm{max}} + i \gamma_{\mathrm{max}}$ and the associated $\mathbf{k_{\mathrm{max}}}$ for the Most Unstable Mode (MUM) of a particular VDF. 
To fully describe every MUM, we use the derived solution for ($\omega_{\mathrm{max}}, \mathbf{k_{\mathrm{max}}}$) provided by \texttt{PLUMAGE} as an input to \texttt{PLUME} solver, yielding
the power emitted for each component over the course of a single wave period $P_j$, electric $\mathbf{\delta E}$ and magnetic $\mathbf{\delta B}$ field fluctuations and polarizations, as well as component density $\delta n_j$ and velocity $\mathbf{\delta v_j}$ fluctuations corresponding to $\gamma_{\mathrm{max}}$. 

\section{Derived Instability thresholds}
\label{sec:si.thresholds}

In order to extract parametric limits, we use \texttt{PLUMAGE} to scan over a broad parameter space where the ODI is expected. 
We cover core proton and isotropic $\alpha$-particle plasmas spanning $\beta_{\parallel,p} \in [10^{-3}, 10^{-1}]$, $T_{\perp,p}/T_{\parallel,p} \in [1, 30]$, $n_\alpha/n_p \in [0, 0.075]$, and $\Delta v_{\alpha}/v_A \in [0, 1.3]$. 

\begin{figure}
\includegraphics[width=0.47\textwidth]{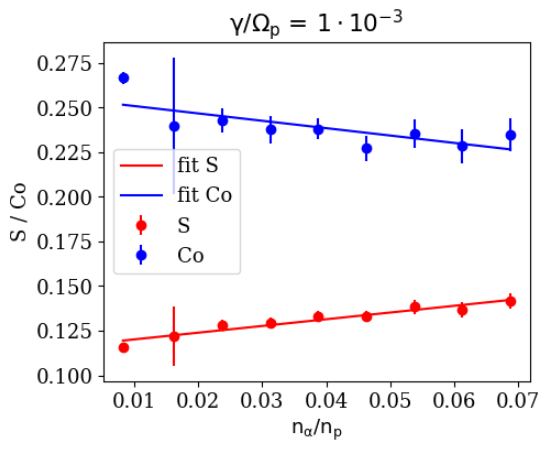}
\caption{The values of slopes and cutoffs derived as shown in Fig. \ref{fig:beta_vva_threshold} of the main article, for various values of $n_\alpha/n_p$. 
Lines show fitted linear trends to be used in Equation \ref{eq:supp_thresholds}. 
}
\label{fig:na_fit}
\end{figure}

ODI thresholds show a remarkably linear trend in logarithmic $\beta_{\parallel,p} - \Delta v_\alpha / v_A$ space. 
We are therefore able to perform accurate linear fits as shown on Fig. \ref{fig:beta_vva_threshold} of the main article. 
Every fit performed in such a way is valid for a single relative density of $\alpha$-particles $n_\alpha/n_p$. 
Plotting these results with respect to $n_\alpha/n_p$, it is visible that the fitted slopes and cutoffs themselves exhibit linear trends (Fig. \ref{fig:na_fit}). 
This result provides a simple numerical approximation for the ODI threshold in $\Delta v_\alpha/v_A$ for various growth rates as a function of $\beta_{\parallel,p}$ and $n_\alpha/n_p$.
Although occasional outliers are present, the linear trends are consistent throughout a very large phase space, spanning over the relative densities and growth rates relevant for solar wind conditions. 

Every linear fit in $\beta_{\parallel,p} - \Delta v_\alpha / v_A$ space for given $n_\alpha / n_p$ provides a slope and an intercept, both of which are single data points as shown on Fig. \ref{fig:na_fit}. 
Then, another linear fit can be made for all the available relative $\alpha$-particle densities, allowing for an analytical expression in the form 
\begin{equation}
    \Delta v_\alpha / v_A (\gamma / \Omega_p) \geq 10^{b_{Co} + a_{Co} n_\alpha/n_p} \beta_{\parallel,p}^{b_{S} + a_{S} n_\alpha/n_p}
    \label{eq:supp_thresholds}
\end{equation}
that gives the ODI threshold for a given growth rate $\gamma / \Omega_p$. 
We summarize all the results of these fits in Table \ref{tab:thresholds_alpha}.

\begin{table}
    \centering
    \begin{tabular}{| c | c | c | c | c |}
    \hline
    $\gamma/\Omega_p$ & a$_S$ & b$_S$ & a$_{Co}$ & b$_{Co}$ \\
    \hline
    $1 \cdot 10^{-4}$ & 0.09 ± 0.05 & 0.133 ± 0.002 & -0.6 ± 0.2 & 0.258 ± 0.007 \\
    $2 \cdot 10^{-4}$ & 0.19 ± 0.03 & 0.129 ± 0.001 & -0.6 ± 0.2 & 0.26 ± 0.007 \\
    $5 \cdot 10^{-4}$ & 0.26 ± 0.03 & 0.125 ± 0.001 & -0.5 ± 0.2 & 0.258 ± 0.007 \\
    $1 \cdot 10^{-3}$ & 0.38 ± 0.04 & 0.116 ± 0.002 & -0.4 ± 0.1 & 0.255 ± 0.006 \\
    $2 \cdot 10^{-3}$ & 0.43 ± 0.05 & 0.107 ± 0.002 & -0.7 ± 0.2 & 0.268 ± 0.009 \\
    $5 \cdot 10^{-3}$ & 0.35 ± 0.05 & 0.1 ± 0.002 & -1.3 ± 0.3 & 0.3 ± 0.01 \\
    $1 \cdot 10^{-2}$ & 0.46 ± 0.06 & 0.083 ± 0.003 & -0.7 ± 0.2 & 0.268 ± 0.009 \\
    $2 \cdot 10^{-2}$ & 0.1 ± 0.2 & 0.07 ± 0.01 & -1.4 ± 0.8 & 0.29 ± 0.04 \\
    \hline
    \end{tabular}
    \caption{Table of fitted instability thresholds for various ODI growth rates for proton and $\alpha$-particle plasma, with values of coefficients used in Equation \ref{eq:supp_thresholds}. }
    \label{tab:thresholds_alpha}
\end{table}

We repeat the same process for a plasma consisting of proton core and proton beam particles. 
The parameter space we investigate is the same for $\beta_{\parallel,c}$ and $T_{\perp,c}/T_{\parallel,c}$, while we expand $n_b/n_c \in [0, 0.3]$, and $\Delta v_b/v_A \in [0, 2.5]$ to ensure that all the same mass fluxes as the $\alpha$ case are captured. 
The two sets of \texttt{PLUMAGE} runs contain a total of 726,800 parameter combinations. 
The results are summarized in Table \ref{tab:thresholds_beam}. 
The trends from Fig. \ref{fig:numerical_thresholds}---which shows thresholds for $\gamma / \Omega_p \geq 10^{-3}$---are maintained for a wide range of growth rates. 
For the same mass flux, ODI requires marginally lower drifts from $\alpha$-particles than from proton beams to activate. 

\begin{table}
    \centering
    \begin{tabular}{| c | c | c | c | c |}
    \hline
    $\gamma/\Omega_p$ & a$_S$ & b$_S$ & a$_{Co}$ & b$_{Co}$ \\
    \hline
    $1 \cdot 10^{-4}$ & -0.01 ± 0.04 & 0.152 ± 0.003 & -0.61 ± 0.1 & 0.43 ± 0.006 \\
    $2 \cdot 10^{-4}$ & -0.08 ± 0.03 & 0.157 ± 0.002 & -0.64 ± 0.05 & 0.433 ± 0.005 \\
    $5 \cdot 10^{-4}$ & -0.08 ± 0.01 & 0.157 ± 0.001 & -0.56 ± 0.04 & 0.428 ± 0.005 \\
    $1 \cdot 10^{-3}$ & -0.08 ± 0.01 & 0.156 ± 0.001 & -0.59 ± 0.03 & 0.43 ± 0.004 \\
    $2 \cdot 10^{-3}$ & -0.06 ± 0.01 & 0.152 ± 0.002 & -0.5 ± 0.02 & 0.422 ± 0.004 \\
    $5 \cdot 10^{-3}$ & -0.02 ± 0.01 & 0.141 ± 0.002 & -0.47 ± 0.03 & 0.415 ± 0.005 \\
    $1 \cdot 10^{-2}$ & 0.0 ± 0.02 & 0.133 ± 0.003 & -0.52 ± 0.04 & 0.423 ± 0.006 \\
    $2 \cdot 10^{-2}$ & 0.07 ± 0.02 & 0.106 ± 0.003 & -0.4 ± 0.05 & 0.391 ± 0.008 \\
    \hline
    \end{tabular}
    \caption{Table of fitted instability thresholds for various ODI growth rates for proton core and proton beam plasma, with values of coefficients used in Equation \ref{eq:supp_thresholds}. }
    \label{tab:thresholds_beam}
\end{table}

\section{Hybrid Simulation Details}
\label{ssec:si.simulation}

We use recently developed parallelized 2.5D and 3D hybrid codes to model multi-ion, magnetized solar wind plasma. These models build on the 1D hybrid formulation \cite{Winske_1993}, later extended to 2D \cite{Ofman_2007_JGRA} and parallelized \cite{Ofman_2010_JGRA}. The hybrid approach captures ion dynamics across many ion--cyclotron periods.

In the 2.5D hybrid code, we evolve three velocity components and three field components in two spatial dimensions. Recent studies apply a $256^2$ grid with up to 512 particles per cell, effectively capturing plasma dispersion and evolution. Higher resolutions such as $512^2$ or $1024^2$ can also be used depending on the simulation requirements. Each super-particle is represented by a Gaussian distribution across multiple cells, approximating many physical particles. In warm, multi-ion plasma, the ion kinetic dissipation scale sets the smallest length scales that must be resolved.

Each numerical particle represents a large number of real particles, based on density normalization. The number of particles per cell is chosen to control statistical noise and ensure accurate VDF resolution, which generally depends on the plasma $\beta$. Numerical stability and energy conservation are verified through convergence tests, ensuring that numerical noise remains below physical fluctuation levels.

The following equations of motion are solved for each particle of species $k$:
\begin{equation}
\frac{d \vec{x}_k}{dt} = \vec{v}_k, \qquad  
m_k \frac{d \vec{v}_k}{dt} = Ze\left(\vec{E} + \vec{v}_k \times \vec{B}\right),
\end{equation}
where $m_k$ is the ion mass, $Z$ is the charge number, $e$ is the electron charge, and $c$ is the speed of light. Neglecting electron inertia allows the electron momentum equation to be solved for the electric field, yielding a generalized Ohm’s law:
\begin{equation}
0 = e n_e\left(\vec{E} + \vec{v}_e \times \vec{B}\right) + \nabla p_e,
\end{equation}
where $p_e = n_e k_b T_e$, with either isothermal or adiabatic electrons used for closure, and quasi-neutrality requires $n_e = n_p + Z n_i$.

Electromagnetic fields are computed on a 2D or 3D spatial grid, and particle equations of motion are solved as protons and ions respond to the fields at each time step. High-order filtering terms are applied as needed to reduce numerical noise. This modeling framework is validated in multiple solar wind studies \cite{Ofman_2010_JGRA, Ofman_2022_ApJ,Ofman_2025_ApJ, Yogesh_2025_ApJ}. The particle and field equations are integrated using the Rational Runge--Kutta method \cite{wambecq1978}, and spatial derivatives are computed using a pseudo-spectral FFT method for periodic boundaries or finite-difference operators for non-periodic boundaries \cite{Ofman_2013_JGRA}.

The hybrid model computes the self-consistent evolution of ion VDFs, including nonlinear wave--particle interactions. 
Recent 3D hybrid simulations model solar wind plasma dynamics and wave--particle interactions \cite{Vasquez2015, Ofman_2019_SolPhys, Ofman_2022_ApJ}. The model captures inhomogeneous structures and waves, including oblique and kinetic Alfv\'{e}n waves \cite{Ofman2017}. Unstable VDFs---such as drifting or anisotropic ion populations---generate a self-consistent wave spectrum. Super-Alfv\'{e}nic drifts drive drift instabilities that produce non-Maxwellian distributions in the nearly collisionless solar wind \cite{Xie2004, Ofman2007}, while bi-Maxwellian conditions can similarly yield non-Maxwellian VDFs \cite{Ofman2014, Ofman2017}.

\begin{figure*}
    \centering
    \animategraphics[loop,autoplay,width=0.6\linewidth]{3}{PRL_ALPS_Video_Frames/frame_}{000}{049}
    \caption{The evolution of the ion VDFs in a hybrid simulation (\emph{top}) shown in animated gif with analysis of the emission and absorption of the oblique, forward propagating \Alfven wave. 
    The information on the panels is as: 
    (top) Proton and alpha VDFs, along with associated $v_{\textrm{res},j}^n/v_A$; 
    (second) $v_{\textrm{res},j}^n (|k|d_p)$; 
    (third) $|\gamma|/\Omega_p (|k|d_p)$; 
    (fourth) $|\gamma_p|/\omega_r (|k|d_p)$ and $|\gamma_p^{n=1}|/\omega_r (|k|d_p)$; 
    (fifth) $|\gamma_\alpha|/\omega_r (|k|d_p)$ and $|\gamma_\alpha^{n=0,-1}|/\omega_r (|k|d_p)$; 
    (bottom two rows) $\mathcal{G}[f_j(v_\perp/v_A,|k|d_p)]$ highlighting velocities responsible for growth or damping. 
    The gif file with various times in simulation  is available on opening file with Acrobat viewer. 
    }
    \label{fig:hyb2d_ev}
\end{figure*}


\section{Numerical Evaluation of the Linear Stability Hybrid Simulations}
\label{sec:si.alps}

Rather than calculating the linear plasma response by modeling the ion VDFs with one or two bi-Maxwellian distributions, we use the Arbitrary Linear Plasma Solver (\texttt{ALPS}) \cite{Verscharen_2018_JPlPh,Klein_2026_GRL} to determine the normal modes of the system without assuming an analytic form for the VDFs.
\texttt{ALPS} is a parallelised numerical code that solves the Vlasov-Maxwell dispersion relation in hot magnetised plasma, allowing for any number of particle species with arbitrary gyrotropic VDFs, $f_j(v_\perp,v_\parallel)$, supporting waves with any direction of propagation with respect to the background magnetic field. 
The wave equation is solved by numerically integrating the dispersion relation integrals that are inputs for the elements of the species susceptibility tensor $\underline{\underline{\chi}}_j$, which in turn defines the dielectric tensor 
\begin{equation}
\underline{\underline{\epsilon}}(\omega,\mathbf{k})= \underline{\underline{1}}
+ \sum_j \underline{\underline{\chi}}_j(\omega,\mathbf{k}),
\label{eqn:dielectric}
\end{equation}
where $\mathbf{n}=c \mathbf{k}/\omega$ is the complex index of refraction, with $z$ oriented in the direction of the background magnetic field.
For a selected wavevector $(k_\perp,k_\parallel)d_p$, we determine complex frequency normal mode $(\omega_r,\gamma)/\Omega_p$ by identifying non-trivial solutions of the homogeneous wave equation
\begin{align}
\underline{\underline{\Lambda}} \cdot \mathbf{E} = & \nonumber
\left(
\begin{array}{ccc}
    \epsilon_{xx} - n_z^2 & \epsilon_{xy} & \epsilon_{xz} + n_x n_z \\
   \epsilon_{yx} &  \epsilon_{yy} - n_x^2 - n_z^2& \epsilon_{yz} \\
    \epsilon_{zx }+ n_x n_z  &  \epsilon_{zy}  & \epsilon_{zz} - n_x^2 
\end{array}
\right) 
\left(
\begin{array}{c}
   E_x \\ E_y \\ E_z 
\end{array}
\right) \\
&=0. 
\label{eqn:wave_equation}
\end{align}
Additional details on the equations and numerical methods can be found in \citet{Klein_2025_PhPl}.

To determine the linear response of the proton and alpha VDFs from the hybrid simulation, for each time step we first gyrotropize the volume-averaged, magnetic-field-aligned distributions,
\begin{equation}
f_j(v_\perp,v_\parallel)=2 \pi \int d\theta v_\perp f(v_\parallel,v_{\perp,1},v_{\perp,2}).
\end{equation}
\texttt{ALPS} normalizes velocity dimensions by the \Alfven velocity, computed using the mass density of the reference species, $v_{A,ref}$, which for this calculation we select to be the protons.

Given the low-$\beta$ values of these simulations, the ion VDFs are very narrow in these units.
In order to provide a sufficiently wide range of velocities with non-zero phase space density, we follow the procedure described in \cite{Klein_2026_GRL}, where a two-component bi-Maxwellian fit is performed for each ion VDF.
These fits are evaluated over a collar of width $v_{A,p}$ to smoothly extend the simulated distribution and thus remove numerical discontinuities as resonant velocities pass over the edge of the simulated VDF.
We next apply a thin-plate spline to the  combined gyrotropized VDF and collar, creating a smooth Cartesian representation on which velocity gradients $\partial v_\perp$ and $\partial v_\parallel$ can be calculated. 
We choose a grid of $(n_\perp\times n_\parallel) = (200 \times 401)$ points; increasing the grid resolution, not shown, does not quantitatively impact the resulting dispersion relations.
We treat the electrons as an isotropic background distribution, with $T_e=T_{\parallel,p}$, with density and drift necessary to enforce quasi-neutrality and zero net current.

We identify the forward-propagating \Alfven solution, and then using a secant method, follow that solution for varying wavevectors, illustrated in Figs.~\ref{fig:double-scan} and \ref{fig:GQL}.
In addition to the complex frequency, we determine the eigenfunctions for the perturbed densities $\delta n_s$, velocities $\delta \mathbf{U}_s$, and electromagnetic fields $\mathbf{E}$ and $\mathbf{B}$ by evaluating the linearized Maxwell's equations, the continuity equation, and the wave equation (see \cite{Klein_2025_PhPl} for details). 
The electric fields can be combined with the susceptibilities for each component to compute the power emitted or absorbed by each component in the weak damping limit $|\gamma|\ll |\omega_r|$, 
\begin{equation}
    \frac{\gamma_j(\mathbf{k})}{\omega_{\text{r}}(\mathbf{k})} = \frac{\mathbf{E}^*(\mathbf{k}) \cdot \underline{\underline{\chi}}_j^a(\mathbf{k}) \cdot \mathbf{E}(\mathbf{k})}{4 W_{\text{EM}}(\mathbf{k})}.
\end{equation}
Here, $\underline{\underline{\chi}}_j^a(\mathbf{k})$ represents the anti-Hermitian component of the susceptibility for component $j$ evaluated at $\gamma=0$, $\mathbf{E}^*$ represents the complex conjugate of the fluctuating electric field, and 
\begin{equation}
W_{\text{EM}} = \mathbf{B}^*(\mathbf{k})\cdot\mathbf{B}(\mathbf{k}) + \mathbf{E}^*(\mathbf{k})\cdot \frac{\partial}{\partial \omega}[\omega \epsilon_h(\mathbf{k})]\cdot\mathbf{E}(\mathbf{k})
\end{equation}
is the electromagnetic wave energy, where $\epsilon_h$ is the Hermitian part of the dielectric tensor.
We can further decompose the damping in contributions from each cyclotron harmonic, following \cite{Huang_2024_JPlPh},
\begin{multline}
P^{LD,n=0}_s =  \frac{i \omega}{16 \pi} \left[ (\chi_{zz,s}^{(n=0)}-\chi_{zz,s}^{(n=0)*})E_{z}E_{z}^* \right. \\
\left. +\chi_{zy,s}^{(n=0)}E_{y}E_{z}^*-\chi_{zy,s}^{(n=0)*}E_{y}^*E_{z})\right]_{\omega=\omega_{\rm r}}, 
\label{eqn:power_ld.hp}
\end{multline}
\begin{multline}
P^{TTD,n=0}_{s} =  \frac{i \omega}{16 \pi} \left[ (\chi_{yy,s}^{(n=0)}-\chi_{yy,s}^{(n=0)*})E_{y}E_{y}^*\right.  \\
 \left.+\chi_{yz,s}^{(n=0)}E_{y}E_{z}^*-\chi_{yz,s}^{(n=0)*}E_{y}^*E_{z})\right]_{\omega=\omega_{\rm r}}. 
\label{eqn:power_ttd.hp}
\end{multline}
and
\begin{multline}
P_s^{n\neq 0}=  \frac{\omega}{8 \pi}
\left[|E_{x}|^2\left(\chi_{xx,s}^{n} -\chi_{xx,s}^{n,*}\right) 
+ |E_{y}|^2\left(\chi_{yy,s}^{n} -\chi_{yy,s}^{n,*}\right) \right. \\
\left.+  \left(E_{x}^*E_{y} - E_{y}^*E_{x}\right)\left(\chi_{xy,s}^{n} -\chi_{yx,s}^{n,*}\right)
\right]_{\omega=\omega_{\rm r}}. 
\label{eqn:power_cd.hp}
\end{multline}

\subsection{Evolution of Instabilities}
\label{ssec:si.interp}

\begin{figure*}
    \centering
    \includegraphics[width=0.95\linewidth]{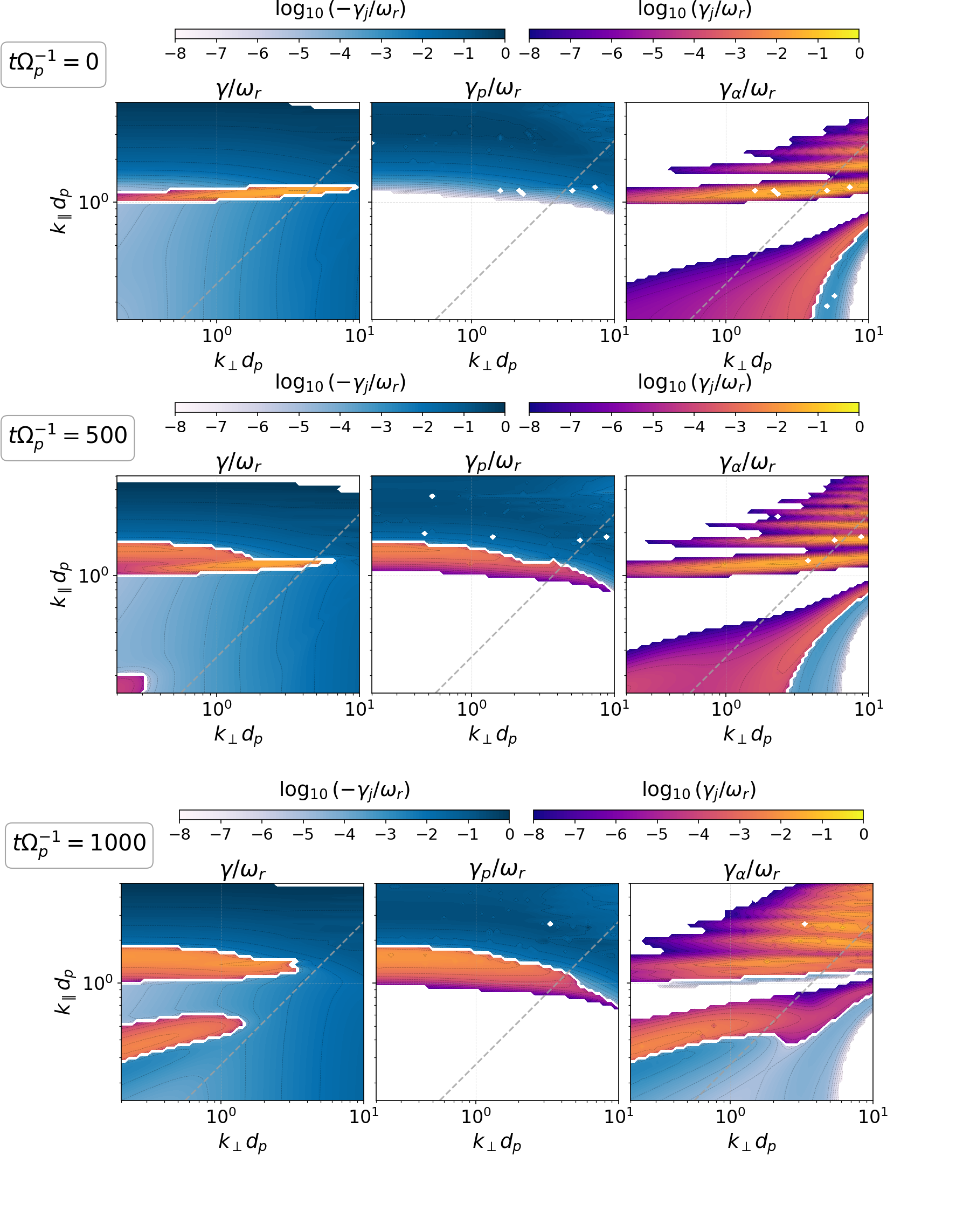}
    \caption{Instability evolution for the forward propagating \Alfven wave surface from simulation initialization (top row) to intermediate and later times (middle and bottom).
    The left column illustrates the total growth or damping $\gamma/\omega_r$, and the middle and right columns the proton and alpha contributions, $\gamma_p/\omega_r$ and $\gamma_\alpha/\omega_r$.
    The grey line indicates the angle of maximum instability at the onset of the simulation.}
    \label{fig:double-scan}
\end{figure*}

To identify which structures in $f_j(v_\perp,v_\parallel)$ contribute to a wave's stability or instability, we apply quasilinear theory \cite{Kennel_1966_PhFl,Verscharen_2013_ApJ_Alpha}
modeling resonant wave-particle interactions by applying the operator
\begin{equation}
\mathcal{G} \equiv \left(1-\frac{k_\parallel v_\parallel}{\omega_{\textrm{r}}} \right)
\frac{\partial}{\partial v_\perp}+
\frac{k_\parallel v_\perp}{\omega_{\textrm{r}}}
\frac{\partial}{\partial v_\parallel}
\label{eqn:ql}
\end{equation}
to $f_p(v_\perp,v_\parallel)$.
Specifically, we determine the VDF regions driving wave growth and damping as a function of scale (\emph{bottom} panels on Fig. \ref{fig:hyb2d_ev}) by constructing functions that depend on $k_\parallel$ and $k_\perp$, applying $\mathcal{G}$ to $f_p$ and $f_\alpha$ across $v_\perp$ at the resonant parallel velocity associated with each wavevector,
\begin{equation}
\frac{v_\parallel^{\textrm{res}}(\mathbf{k})}{v_A}=\frac{\omega_{\textrm{r}}(\mathbf{k})/\Omega_p-n}{k_\parallel d_p}.
  \label{eqn:Gfv}
\end{equation}
The resonant integer $n$ determines the order of the relevant resonance, which is determined by the polarization of the wave. 

\begin{figure*}
    \centering
    \includegraphics[width=0.75\linewidth]{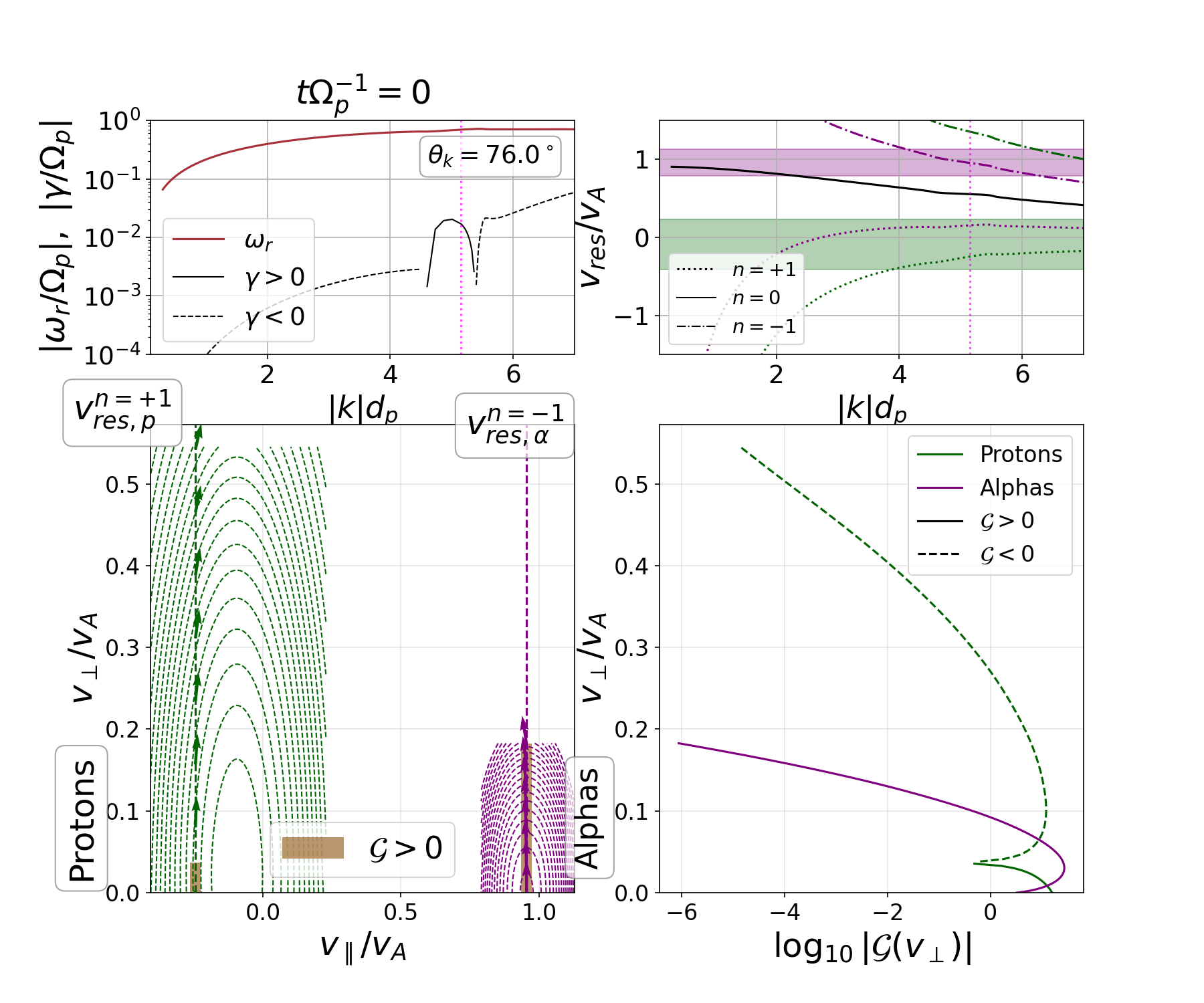}\\
        \includegraphics[width=0.75\linewidth]{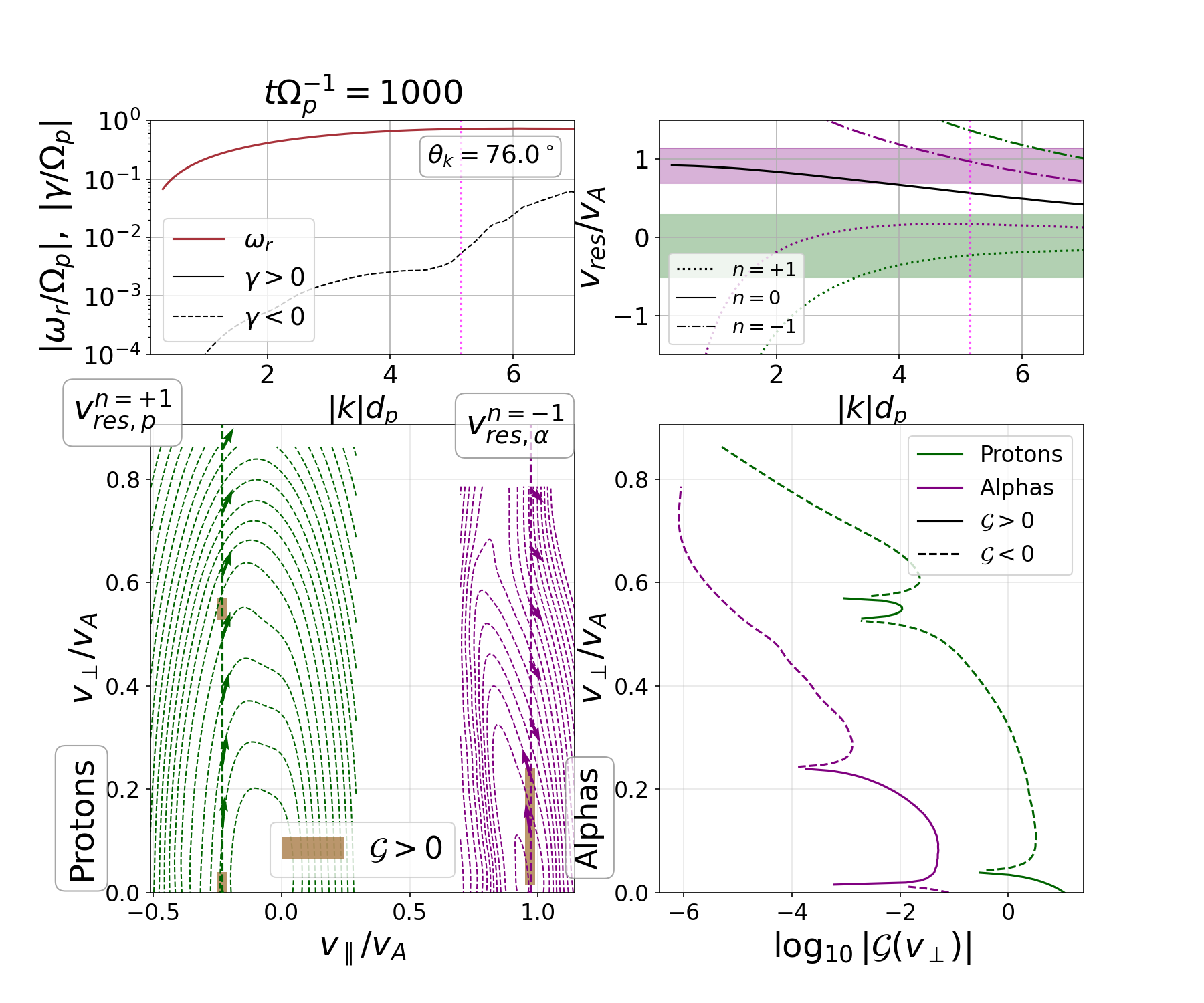}
    \caption{Resonant cyclotron coupling with the protons (green, $n=1$) and alphas (purple, $n=-1$) at the start of the simulation (left) and at $t\Omega_p^{-1}=1000$ (right) for the forward \Alfven solution with $\theta_{kB}=76^\circ$.
    (top left) real frequency $\omega_r/\Omega_p$ (red) and damping/growth rate (dashed and solid black). 
    (top right) $v_{\textrm{res}}^n$ for $n=0,\pm1$, with regions of $v_\parallel$ resolved by the simulation shaded and selected resonant $|k|d_p$ shown in magenta.
    (bottom left) Contours of $f_p(v_\perp,v_\parallel)$ and $f_\alpha(v_\perp,v_\parallel)$ along with resonant velocities. Arrows indicate direction of quasilinear scattering, with brown highlighting regions driving wave growth.
    (bottom right) Velocity resolved quasilinear operator $\mathcal{G}$, Eqn.~\ref{eqn:ql} illustrating regions of emission (solid lines) and damping (dashed).
    }
    \label{fig:GQL}
\end{figure*}

The amount of energy gained or lost is proportional to $\mathcal{G}\{f_p[v_\parallel^{\textrm{res}}(k_\parallel),v_\perp)]\}$, shown for a selected wavevector for two times from the simulation in Fig.~\ref{fig:GQL}.
The diffusive flux of particles flows tangent to semicircles with $v_\perp^2 + (v_\parallel-\frac{\omega_r}{k_\parallel})^2=$ constant, so the sign of the energy transfer depends if the particles gain or lose kinetic energy $v_\perp^2 + v_\parallel^2$.
For visual clarity, the $v_\perp$ values for the selected $v_{\textrm{res}}$ that are emitting power are highlighted in brown.
For these oblique wavevectors, we see that at early times, the entire simulated $\alpha$-particle distribution contributes to driving the instability through the $n=-1$ alpha cyclotron resonance.
At the same time, the proton VDF is heated through the $n=+1$ proton cyclotron resonance.
As the protons do not absorb as much energy as the alphas emit, the equilibrium is unstable and a coherent wave is observed.
Both the proton and alpha VDFs increase their perpendicular kinetic energy.
In Fig.~\ref{fig:double-scan}, we illustrate the wavevector regions of stability and instability for the forward-propagating \Alfven dispersion surface, $\gamma/\omega_r$, as well as the power being emitted or absorbed by the protons $\gamma_p/\omega_r$, and the $\gamma_\alpha/\omega_r$.
The MUM is at $\theta_{kB}\sim 75^\circ$ and $k_\parallel d_p \sim 1$.
It is visible that at early times, the protons do not contribute to any wave growth, but strongly damp via the $n=1$ resonance.

\begin{figure*}
    \centering
    \includegraphics[width=0.95\linewidth]{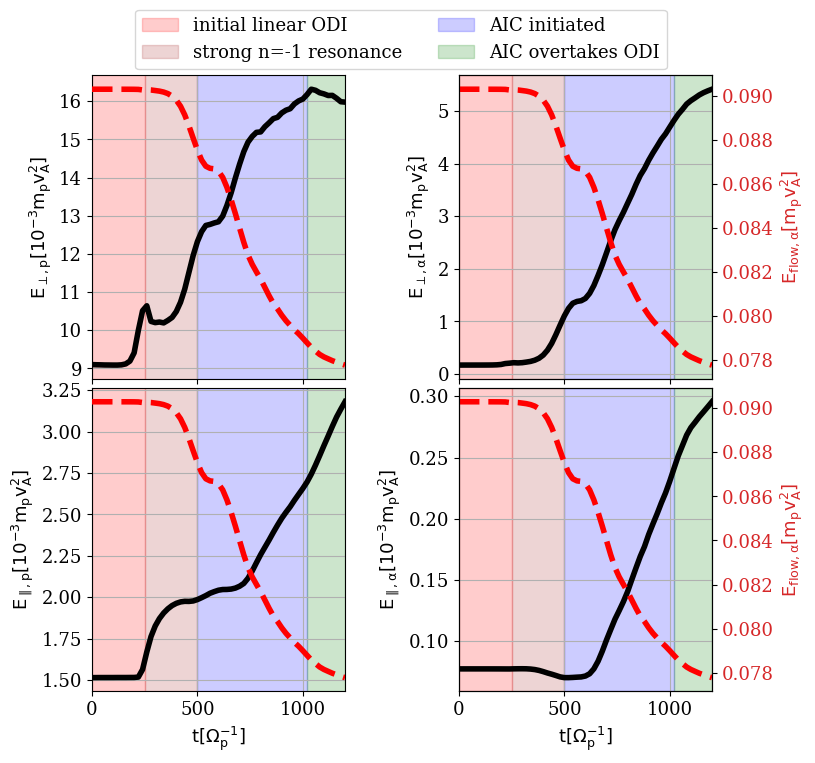}
    \caption{Time series of the VDF moments computed with the 2.5D hybrid model. 
    All values are normalized to the units of $m_p v_A^2$ to illustrate the transfer of $\alpha$-particle drift energy to both proton and $\alpha$-particle thermal energy in both parallel and perpendicular direction. The parallel $(E_{||})$ and perpendicular $(E_{\perp})$ thermal energies of protons and $\alpha$ particles are shown in black, the kinetic energies due to the drift are shown in red dashes. 
    All trends are monotonic except near transition between regimes marked with colored background due to highly irregular VDF shapes near transition times. }
    \label{fig:time_series}
\end{figure*}

The interaction between the two instabilities is depicted on Fig. \ref{fig:time_series}. 
After the initial scattering of smaller number of particles by linear ODI, a larger fraction of the VDF becomes susceptible to resonance, which causes intense heating at the expense of the $\alpha$-particle drift. 
By $t\Omega_p=500$, this strong cyclotron damping has elongated the proton VDF in the perpendicular direction sufficiently so that the parallel \Alfven cyclotron waves are unstable.
There is also a large scale instability driven by the $n=0$ resonance with the alpha distribution, and the $n=-1$ resonance is still driving an oblique instability. 
Hence, during the period marked in \emph{blue} on Figure \ref{fig:time_series} we observe steady transfer of drift energy into three distinct processes: perpendicular heating, which is partially mitigated by AIC for protons, and parallel heating \textbf{via $n=0$ resonance}. 

At a later time in the simulation, $t\Omega_p^{-1}=1000$, the ion VDFs have been sufficiently deformed to alter their linear response. 
As the drift energy drains, the overall equilibrium is stable at this oblique angle, though two parallel instabilities are now active---the parallel proton-driven $n=1$ instability at $k_\parallel d_p \sim 1$ is the fastest growing mode, followed by a parallel propagating, larger scale $n=0$ alpha-driven instability. 
A narrower region of the alpha VDF with $v_\perp<0.2 v_A$ emits power, but due to the heating of the distribution, there are fewer particles at these velocities and the overall power emitted is less than at $t\Omega_p^{-1}=0$.
The protons absorb all of the power emitted by the alphas, continuing to heat that population. 
As a result, the oblique $n=-1$ alpha driven instability is nearly extinguished.

A more detailed exploration of the VDF response for the oblique \Alfven instability is shown on various frames of Fig. \ref{fig:hyb2d_ev}. 
Here, we briefly describe the same time frames as shown in Fig.~\ref{fig:GQL}.
Rather than focusing on a single wavevector, we investigate the emission and absorption from the $n=\pm1$ and $0$ resonances and the quasilinear diffusion operator $\mathcal{G}$ for both species as a function of wavevector.

At $t\Omega_p^{-1}=0$, we see that the instability is entirely driven by the $n=-1$ alpha-cyclotron resonance, with significant absorption from the $n=1$ proton-cyclotron resonance.
No other resonances significantly contribute to the wave behavior across all scales for this propagation angle.
The alpha-driven instability is tightly localized in wavevector space and limited to low $v_\perp/v_A$ values.

At $t\Omega_p^{-1}=1000$, the region of emitting alphas has expanded in both wavevector and $v_\perp$, with higher (lower) $v_\perp$ particles emitting at larger (smaller) $k d_p$ values.
We see that for even smaller wavevectors, the alpha VDF couples to with the $n=0$ resonance, which at less oblique angles, drives a parallel cyclotron instability at $k_\parallel d_\alpha \sim 1$, distinct from the smaller scale proton driven parallel cyclotron instability.
At oblique angles, the protons continue to absorb via the $n=1$ resonance, continuing to heat the protons to higher $T_\perp$.

